\documentclass[nofootinbib,preprint,superscriptaddress,showpacs,amsmath,amssymb,aps,pra,showkeys]{revtex4-1}

\usepackage{amsmath,amsfonts,amssymb}
\usepackage{graphicx}
\usepackage{color}

\begin{document}

\title{Multielectron polarization effects in strong-field ionization: Narrowing of momentum distributions and imprints in interference structures}

\author{N. I. Shvetsov-Shilovski}
\email{n79@narod.ru}
\affiliation{Institut f\"{u}r Theoretische Physik, Leibniz Universit\"{a}t Hannover, D-30167 Hannover, Germany}

\author{M. Lein}
\affiliation{Institut f\"{u}r Theoretische Physik and Centre for Quantum Engineering and Space-Time Research, Leibniz Universit\"{a}t Hannover, D-30167 Hannover, Germany}

\author{L. B. Madsen}
\affiliation{Department of Physics and Astronomy, Aarhus University, 8000 {\AA}rhus C, Denmark}

\date{\today}

\begin{abstract}
We extend the semiclassical two-step model [Phys.~Rev.~A~\textbf{94}, 013415 (2016)] to include a multielectron polarization-induced dipole potential. Using this model we investigate the imprints of multielectron effects in the momentum distributions of photoelectrons ionized by a linearly polarized laser pulse. We predict narrowing of the longitudinal momentum distributions due to electron focusing by the induced dipole potential. We show that the polarization of the core also modifies interference structures in the photoelectron momentum distributions. Specifically, the number of fanlike interference structures in the low-energy part of the electron momentum distribution may be altered. We analyze the mechanisms underlying this interference effect. The account of the multielectron dipole potential seems to improve the agreement between theory and experiment.\\

 
\end{abstract}


\maketitle

\section{Introduction}

Advances in laser technologies, especially the advent of table-top intense femtosecond optical laser systems have led to the remarkable progress in strong-field physics that studies the interaction of strong laser radiation with atoms and molecules. This interaction results in such phenomena as above-threshold ionization (ATI), high-order harmonic generation (HHG), nonsequential double ionization (NSDI), etc. (see, Refs.~\cite{BeckerRev2002, MilosevicRev2003, FaisalRev2005, FariaRev2011} for reviews). In atomic ATI an electron absorbs more photons than necessary for ionization. The kinetic-energy spectrum generated by the ATI process consists of two distinct parts: a rapidly decaying low-energy part of the spectrum that ends at an energy around $2U_{p}$, where $U_p=F^2/4\omega^2$ is the ponderomotive energy (atomic units are used throughout the paper unless indicated otherwise); this part is followed by the high-energy plateau extending up to $\sim10U_{p}$, which is often several orders of magnitude less intense than the maximum of the low-energy part. The part of the spectrum below $2U_{p}$ is mainly formed by electrons that do not undergo hard recollisions with their parent ions. These electrons are usually referred to as direct electrons. The spectrum of the direct electrons can be described by the two-step model for ionization \cite{Linden88, Gallagher88, Corkum89}. In the first step of this model an electron is promoted to the continuum by tunneling ionization \cite{Dau3, PPT, ADK}, and in the second step it moves along a classical trajectory in the laser field. In contrast to this, the high-energy plateau arises due to rescattered electrons that are driven back by the laser field to their parent ions and scatter by large angles. The qualitative description of the rescattering processes is provided by the three-step model \cite{Kulander_Schafer, Corkum1993}, which includes the interaction of the returning electron with the parent ion as the third step. The concept of rescattering also provides the basis of the mechanisms responsible for HHG and NSDI. Indeed, the returning electron can recombine with the residual ion, resulting in emission of high-frequency radiation, or as an alternative, if the energy of the rescattered electron is high enough, it can liberate another electron from the parent ion. 

The main theoretical approaches to strong-field phenomena include the direct numerical solution of the time-dependent Schr\"{o}dinger equation (TDSE) (see, e.g., Refs.~\cite{LaGatutta1990, Cormier1997, Nurhuda1999, Muller1999, Bauer2006, Madsen2007, Grum2010, Patchkovskii2016, Tong2017}), the strong-field approximation (SFA) \cite{Keldysh1964, Faisal1973, Reiss1980, Lewenstein1994}, and semiclassical models employing classical equations of motion to describe the electron motion in the continuum (see, e.g., \cite{Keller2012, Shvetsov2012, Dimitrovski2014, Dimitrovski2015, Dimitrovski2015JPB}). The two-step and the three-step models are the most well-known examples of the semiclassical approaches. All these theoretical methods are usually based on the single-active-electron approximation (SAE). Within the SAE, the ionization is considered as a one-electron process, i.e., an atom (or molecule) in the laser field is replaced by a single electron that interacts with the laser field and an effective potential. The latter is optimized to reproduce the ground state and singly excited states. Although SAE allows an understanding of the major features of ATI and HHG (see, e.g., Refs.~\cite{BeckerRev2002, GrossmannBook}), the role of multielectron (ME) effects in these processes has recently been attracting considerable attention (see recent Refs.~\cite{Kang2018, Le2018} and references therein). 

Among the theoretical approaches capable to account for ME effects in strong-field processes are the time-dependent density-functional theory \cite{RungeGross1984} (see Ref.~\cite{UllrichBook} for a text-book treatment), multiconfiguration time-dependent Hartree-Fock theory~\cite{Zanghellini2004, Caillat2005}, time-dependent restricted-active-space self-consistent-field theory~\cite{Madsen2013}, time-dependent complete-active-space self-consistent-field theory~\cite{Ishikawa2013}, time-dependent $R$-matrix theory~\cite{Burke1997, Lysaght2009}, $R$-matrix method with time-dependence~\cite{Nikolopoulos2008, Moore2011}, time-dependent configuration-interaction-singles~\cite{Rohringer2006, Pabst2012}, time-dependent restricted-active space configuration-interaction methods~\cite{Hochstuhl2012, Bauch2014}, time-dependent analytical $R$-matrix theory~\cite{Torlina2012}, and various semiclassical models (see, e.g., Refs.~\cite{Keller2012, Shvetsov2012, Dimitrovski2014, Dimitrovski2015, Dimitrovski2015JPB, Kang2018}). The advantages of the semiclassical approaches, such as their relative numerical simplicity and the ability to provide an illustrative physical picture of the phenomena under study, are particularly important in investigations of complex ME dynamics. 

Laser-induced polarization of the ionic core is one of the well-known examples of ME effects. During the last years, significant progress has been achieved in studies of the polarization effects in ATI (see Refs.~\cite{Dimitrovski2010}, \cite{Keller2012, Shvetsov2012, Dimitrovski2014, Dimitrovski2015, Dimitrovski2015JPB} and \cite{Kang2018}). The effective potential for the outer electron that takes into account the laser field, the Coulomb potential as well as the polarization effects of the inner core (see Eq.~(\ref{mepot}) in Sec.~II) was found in Refs.~\cite{Brabec2005, Zhao2007, Dimitrovski2010} within the adiabatic approximation. It was shown that the Schr\"{o}dinger equation with this effective potential and accounting for the Stark shift of the ionization potential can be approximately separated in the parabolic coordinates \cite{Keller2012}. This separation procedure results in a certain tunneling geometry. The corresponding physical picture was named tunnel ionization in parabolic coordinates with induced dipole and Stark shift (TIPIS). The semiclassical model based on the TIPIS approach was validated by comparison with \textit{ab initio} results \cite{Keller2012, Shvetsov2012} and experiments \cite{Keller2012, Dimitrovski2014, Dimitrovski2015, Dimitrovski2015JPB}. It was shown that for different atoms and molecules (Ar, Mg, naphthalene, etc.) the photoelectron momentum distributions are highly sensitive to ME effects as captured by the induced dipole of the atomic core \cite{Keller2012, Shvetsov2012, Dimitrovski2014, Dimitrovski2015, Dimitrovski2015JPB}. 

Most of the studies mentioned here deal with circularly or close to circularly polarized laser fields. The reason is that the potential of Refs.~\cite{Brabec2005, Dimitrovski2010} that is used in semiclassical simulations is valid at large and intermediate distances and not at short distances. It is well known that the rescattering processes are suppressed in close to circularly polarized laser fields \cite{Dietrich1994}, and, therefore, the vast majority of the electron trajectories do not return to the vicinity of the ionic core. However, this is certainly not the case for linearly polarized field. This raises the question regarding the applicability of the TIPIS model for linear polarization of the laser field. As a result, there is a lack of theoretical studies of the ME polarization effects in ATI with linearly polarized field. 

To the best of our knowledge, Ref.~\cite{Kang2018} is the only application of the potential of Refs.~\cite{Brabec2005, Dimitrovski2010} to semiclassical simulations of ATI processes in linearly polarized fields. That study focuses on the modification of the low-energy structures \cite{Blaga2009, Quan2009} and the very low-energy structures \cite{Quan2009, Wu2012} due to polarization effects. To the best of our knowledge, the impact of the polarization of the ionic core on the whole direct part of the spectrum has not been investigated so far. Furthermore, the applicability of the semiclassical model with the potential of Refs.~\cite{Brabec2005, Dimitrovski2010} was not discussed in Ref.~\cite{Kang2018}. Finally, quantum interference was disregarded in all the trajectory-based studies of Refs.~\cite{Keller2012, Shvetsov2012, Dimitrovski2014, Dimitrovski2015, Dimitrovski2015JPB, Kang2018}. Since the ME potential affects both the tunnel exit point and the electron dynamics in the continuum \cite{Shvetsov2012}, an imprint of the polarization effects in the interference patterns of the momentum distributions can be expected.   

In this paper we apply the TIPIS model to ATI and momentum distributions in linearly polarized laser fields and analyze the applicability of this model. In order to study the interference effects due to the polarization of the ionic core, we combine the TIPIS approach with the semiclassical two-step model (SCTS)~\cite{Shvetsov2016}. The SCTS model describes quantum interference and accounts for the ionic potential beyond semiclassical perturbation theory. Recently this model was applied to the study of the intra-half-cycle interference of low energy photoelectrons \cite{Xie2016}, to the analysis of the interference patterns emerging in strong-field photoelectron holography (see Refs.~\cite{Walt2017, Shvetsov2018}), and to the investigation of the subcycle interference upon ionization by counter-rotating two-color fields \cite{Eckart2018}. Using the semiclassical approach we calculate the photoelectron momentum distributions and energy spectra of the ATI in linearly polarized laser field with the account for the ME polarization potential and interference. We then analyze both the dynamic and interference effects induced by the polarization of the ionic core. 

The paper is organized as follows. In Sec.~II we briefly review the TIPIS model, discuss its application to the case of linear polarization, and formulate our new combined model. In Sec.~III we calculate photoelectron momentum distributions and energy spectra, identify the imprints of the ME polarization effect, and reveal by trajectory analysis the physical mechanisms underlying the formation of these imprints. The conclusions are given in Sec.~IV. 

\section{Model}

A detailed derivation of the TIPIS model as well as its applications to simulations of the photoelectron momentum distributions in elliptically polarized fields are presented in Ref.~\cite{Shvetsov2012}. Here we repeat the main points to make the presentation self-contained. We next combine the TIPIS approach with the SCTS model. By doing so we develop a two-step semiclassical model for strong-field ionization with the inclusion of the Stark-shift, the Coulomb potential, and the polarization induced dipole potential, capable of describing quantum interference. 

\subsection{TIPIS model and its application to linearly polarized laser field\textcolor{blue}{s}}

In semiclassical simulations the trajectory of an electron $\vec{r}\left(t\right)$ is calculated using Newton's equation of motion:
\begin{equation}
\frac{d^2\vec{r}}{dt^2}=-\vec{F}\left(t\right)-\nabla V\left(\vec{r},t\right),
\label{newton}
\end{equation} 
where $\vec{F}\left(t\right)$ is the electric field of the laser pulse, and the ionic potential $V\left(\vec{r},t\right)$ is given by:
\begin{equation}
V\left(\vec{r},t\right)=-\frac{Z}{r}-\frac{\alpha_{I}\vec{F}\left(t\right)\cdot\vec{r}}{r^3}.
\label{mepot}
\end{equation}
Here $Z$ is the ion charge. In Eq.~(\ref{mepot}) the ME effect is taken into account through the induced dipole potential $\left[\frac{\alpha_{I}\vec{F}\cdot\vec{r}}{r^3}\right]$, where $\alpha_{I}$ is the static polarizability of the ion. As in Ref.~\cite{Shvetsov2012}, we refer to the second term of Eq.~(\ref{mepot}) as the ME term. It is important to stress that the potential of Eq.~(\ref{mepot}) is valid only at large and intermediate distances (see. Refs.~\cite{Dimitrovski2010, Brabec2005, Zhao2007}). 

In order to integrate Eq.~(\ref{newton}), we need: the starting point of the trajectory and the initial velocity of the electron. To obtain the former, i.e., the tunnel exit point, the approximate separation of the static tunneling problem in parabolic coordinates is used in the TIPIS model. If the static field acts along the $z$-axis, we define the parabolic coordinates as $\xi=r+z$, $\eta=r-z$, and $\phi=\arctan\left(y/x\right)$. Then the approximate separation is valid in the limit $\xi/\eta<<1$ \cite{Keller2012}. The tunnel exit point $z_{e}$ is then found as $z_{e}\approx-\eta_{e}/2$, where $\eta_{e}$ is the solution of the following equation:
\begin{equation}
-\frac{\beta_{2}\left(F\right)}{2\eta}+\frac{m^2-1}{8\eta^2}-\frac{F\eta}{8}+\frac{\alpha_{I}F}{\eta^2}=-\frac{I_p\left(F\right)}{4},
\label{exiteta}
\end{equation}
where $I_{p}\left(F\right)$ is the ionization potential, $m$ is the magnetic quantum number of the initial state, and 
\begin{equation}
\beta_{2}\left(F\right)=Z-\left(1+\left|m\right|\right)\frac{\sqrt{2I_{p}\left(F\right)}}{2}
\label{beta}
\end{equation}
is the separation constant \cite{Shvetsov2012}. The TIPIS model accounts for the Stark shift of the ionization potential:
\begin{equation}
I_{p}\left(F\right)=I_{p}\left(0\right)+\left(\vec{\mu}_{N}-\vec{\mu}_{I}\right)\cdot\vec{F}+\frac{1}{2}\left(\alpha_{N}-\alpha_{I}\right)F^2,
\label{stark}
\end{equation}
where $I_{p}\left(0\right)$ is the field-free ionization potential, $\vec{\mu_{N}}$ and $\vec{\mu}_{I}$ are the dipole moments of an atom (molecule) and of its ion, respectively, and $\alpha_{N}$ is the static polarizability of an atom (molecule). For atoms the term linear in $\vec{F}$ is absent in Eq.~(\ref{stark}). The static field $F$ in Eqs.~(\ref{exiteta}), (\ref{beta}), and (\ref{stark}) should be interpreted as the instantaneous value of the laser field $F\left(t_0\right)$ at the time of ionization denoted by $t_0$. 
 
We assume that the electron starts with zero initial velocity along the direction of the laser field: $v_{0,z}=0$. It can, however, have a nonzero initial velocity $\vec{v}_{0,\perp}$ in the perpendicular direction. The ionization time $t_0$ and the initial transverse velocity $\vec{v}_{0,\perp}$ completely determine the electron trajectory. We distribute $t_0$ and $\vec{v}_{0,\perp}$ according to the static ionization rate \cite{DeloneKrainov1991}:
\begin{equation}  
\label{tunrate}
w\left(t_{0},v_{0, \perp}\right)\sim\exp\left(-\frac{2\kappa^3}{3F\left(t_0\right)}\right)\exp\left(-\frac{\kappa v_{0,\perp}^{2}}{F\left(t_0\right)}\right)
\end{equation} 
with $\kappa=\sqrt{2I_{p}\left(F\right)}$. We omit the preexponential factor in Eq.~(\ref{tunrate}), since for atoms it only slightly affects the shape of the photoelectron momentum distributions that we are interested in.

As the ME term of the potential Eq.~(\ref{mepot}) vanishes at $t\geq t_{f}$, where $t_{f}$ is the time at which the laser pulse terminates, after the end of the pulse an electron moves in the Coulomb field only. The asymptotic momentum of the electron $\vec{k}$ can be found from its momentum $\vec{p}\left(t_f\right)$ and position $\vec{r}\left(t_f\right)$ at the end of the laser pulse (see Refs.~\cite{Shvetsov2009, Shvetsov2012}):
\begin{equation}
\label{mominf}
\vec{k}=k\frac{k\left(\vec{L}\times\vec{a}\right)-\vec{a}}{1+k^2L^2}.
\end{equation}
Here $\vec L=\vec r(t_f)\times\vec p(t_f)$ and $\vec a=\vec p(t_f)\times\vec L - Z\vec r(t_f)/r(t_f)$ are the angular momentum and Runge-Lenz vector, respectively. The magnitude of the asymptotic momentum can be found from energy conservation
\begin{equation}
\frac{k^2}{2}=\frac{p^2(t_f)}{2}-\frac{Z}{r(t_f)}
\label{conserv}
\end{equation}
at the end of the laser pulse. Equipped with the ensemble of $\left(t_0,v_{0,\perp}\right)$, and the corresponding values of the asymptotic momenta, we are now ready to combine the TIPIS approach with the SCTS model. 

\subsection{Combination of the TIPIS approach with the SCTS model}

In order to study the ME polarization-induced interference effects, we combine the TIPIS approach with the SCTS model. In the SCTS model every classical trajectory is associated with \textcolor{blue}{a} phase. The latter is calculated using the semiclassical expression for the matrix element of the quantum mechanical propagator \cite{Miller1971, Walser2003, Spanner2003}. For an arbitrary effective potential $V\left(\vec{r},t\right)$ this phase is given (see Ref.~\cite{Shvetsov2016}):
\begin{equation}
\Phi\left(t_{0},\vec v_0\right)= - \vec v_0\cdot\vec r(t_0) + I_{p}t_{0} - \int_{t_0}^\infty dt\, \left\lbrace\frac{p^2(t)}{2}+V[\vec{r}(t)]-\vec r(t)\cdot\vec\nabla V[\vec{r}(t)]\right\rbrace\ .
\label{Phi_sim}
\end{equation}
If $V\left(\vec{r},t\right)$ is set to the potential of Eq.~(\ref{mepot}), the expression for the phase $\Phi\left(t_{0},\vec v_0\right)$ reads as:
\begin{equation}
\Phi\left(t_{0},\vec v_0\right)= - \vec v_0\cdot\vec r(t_0) + I_{p}t_{0} - \int_{t_0}^\infty dt\, \left\lbrace\frac{p^2(t)}{2}-\frac{2Z}{r}-\frac{3\alpha_{I}\vec{F}\left(t\right)\cdot \vec{r}}{r^3}\right\rbrace\ .
\label{Phi_our}
\end{equation}
For our simulations we use an importance sampling implementation of the SCTS model. In this approach we distribute ionization times $t_0^{j}$ and initial velocities $v_{0}^{j} \left(j=1...n_{p}\right)$ according to the square root of the tunneling probability [Eq.~(\ref{tunrate})]. We solve the equations of motion (\ref{newton}) and find the final (asymptotic) momenta of all $n_p$ trajectories in the ensemble. Then we bin the trajectories in cells in momentum space according to their final momenta. The amplitudes associated with the trajectories reaching the same bin that is centered at a given final momentum $\vec{k}$ are added coherently, and the ionization probability is given by (see Ref.~\cite{Shvetsov2016}):
\begin{equation}
\label{prob1}
\frac{dR}{d^{3}k}=\left|\sum_{j=1}^{n_p}\exp\left[i\Phi\left(t_{0}^{j},\vec v_0^{j}\right)\right]\right|^2. 
\end{equation}
We note that convergence both with respect to the size of the momentum bin and the number of the trajectories must be achieved. The bin size and the number of trajectories in the ensemble needed for convergence strongly depend on the laser-atom parameters. All results provided below have been checked for convergence and the computational parameters are explicitly given in the illustrative examples.  
  
\section{Results and Discussion}

In our simulations we use a few-cycle laser pulse linearly polarized along the $z$ axis and defined in terms of a vector potential:
\begin{equation}
\vec{A}\left(t\right)=\left(-1\right)^{n+1}\frac{F_0}{\omega}\sin^2\left(\frac{\omega t}{2n}\right)\sin\left(\omega t\right)\vec{e}_z,
\label{vecpot}
\end{equation} 
Here $\vec{e}_z$ is a unit vector, $F_0$ is the field strength, $\omega$ is the angular frequency, and $n$ is the number of cycles within the pulse present between $t=0$ and $t=t_{f}$, where $t_{f}=2\pi n/\omega$. The electric field is obtained from Eq.~(\ref{vecpot}) by $\vec{F}\left(t\right)=-\frac{d\vec{A}}{dt}$. We solve the equations of motion (\ref{newton}) using a fourth-order Runge-Kutta method with adaptive stepsize control \cite{Numerical}. 

Here we restrict ourselves to the case of atoms. Apart from the fact that the potential of Eq.~(\ref{mepot}) is inapplicable at small distances, the range of applicability of the TIPIS model is restricted by two conditions (see Ref.~\cite{Shvetsov2012}). First, the field-induced term of Eq.~(\ref{stark}) should not exceed 10\%-20\% of the first term, and this introduces an upper bound for the magnitude of the laser intensity. At the same time, the intensity must not be too low: since in the TIPIS model the ionization probability is described by the tunneling formula [Eq.~(\ref{tunrate})], the Keldysh parameter $\gamma=\omega\kappa/F$ \cite{Keldysh1964} should be less or of the order of one. We also note that the using of static polarizabilities is justified for large wavelengths $\lambda$. The choice of the atomic species and the laser parameters for which (i) the ME effects are more pronounced, and (ii) the TIPIS model is applicable, is thoroughly discussed in Ref.~\cite{Shvetsov2012}. 

We perform our simulations for Mg and Ca. For the Mg atom, $I_p=0.28$~a.u., $\alpha_{N}=71.33$~a.u., and $\alpha_{I}=35.00$~a.u., whereas for Ca $I_p=0.22$~a.u., $\alpha_{N}=169.0$~a.u., and $\alpha_{I}=74.11$~a.u. (see Ref.~\cite{Mitroy2010} for the values of polarizabilities). Note that these atoms have similar ionization potentials, but for Ca the static ionic polarizability that enters the ME term is approximately two times larger than the one for Mg. We do the simulations for the intensities of $3.0\times10^{13}$ W/cm$^2$ (Mg) and $1.0\times10^{13}$ W/cm$^2$ (Ca), and use the wavelength 1600~nm for both atoms. The corresponding Keldysh parameters for Mg and Ca are equal to $\gamma=0.73$ and $\gamma=1.13$, respectively. 

First we analyze the applicability of the TIPIS model to the case of linear polarization. In Ref.~\cite{Kang2018} a cutoff was introduced at a radial distance where the core polarization cancels the laser field. At the distances smaller than the cutoff radius the electron does not experience polarization effects. This approach follows the reasoning of Ref.~\cite{Zhao2007}, which was based on considerations of a behaviour similar to that of a large metallic-like system. 

Here we also introduce a cutoff radius $r_C$. However, in contrast to Ref.~\cite{Kang2018}, we disregard all the trajectories entering the sphere $r\leq r_{C}$. By doing so we prevent the electron trajectories from reaching the vicinity of the residual ion. The elimination of the returning trajectories leads to the depletion of some parts of the photoelectron momentum distributions. It is clear that these depleted parts cannot be reliably calculated within the TIPIS model. However, these domains usually correspond to the upper boundary of the direct ionization spectrum and do not involve its main part containing most of the yield. This point is illustrated by Figs.~1~(a) and (b). In Fig.~1~(a) we show the photoelecton momentum distribution calculated taking into account all the trajectories of the ensemble. The white curve in Fig.~1~(a) shows the boundary of the part of the momentum distribution that is reliably reproduced when the trajectories entering the area $r\leq r_{C}$ are excluded. The electron energy spectra calculated with and without the elimination of the returning trajectories are compared in Fig.~1~(b). If the photoelectron momentum distribution [Eq.~(\ref{prob1})] is available, the energy spectrum can be calculated as follows:
\begin{equation}
\frac{dR}{dE}=2\pi\sqrt{\left(2E\right)}\int_{0}^{\pi}d\theta\sin\theta\frac{dR}{d^{3}k}\left[\vec{k}\left(\theta\right)\right].
\label{spectrum}
\end{equation}

For the parameters of Fig.~1 we need the bin size equal to $6.25\cdot10^{-4}$~a.u. and an ensemble of 3.2 billion trajectories to achieve convergence. The latter was controlled by comparison of the energy spectra within the energy range in which signal decreases to $10^{-5}$ of its maximum. In our simulations we have chosen the cutoff radius equal to $r_{C}=5.0$~a.u., but for the parameters considered here the results only weakly depend on the particular value of $r_C$ in the range from 3.0~a.u to 7.0~a.u. Figure~1~(b) clearly shows that almost the whole direct part of the electron spectrum is unaffected by the exclusion of the returning trajectories. Taking into account these findings, in what follows we do not impose the condition $r<r_{C}$. We note, however, that at different laser-atom parameters the applicability of the TIPIS model to the case of linear polarization may be not as favorable as in Figs.~1~(a) and (b). An analysis similar to the one presented here is, therefore, needed for any set of laser parameters before application of the TIPIS model to linearly polarized fields. 

\begin{figure}[h]
\begin{center}
\includegraphics[width=.8\textwidth]{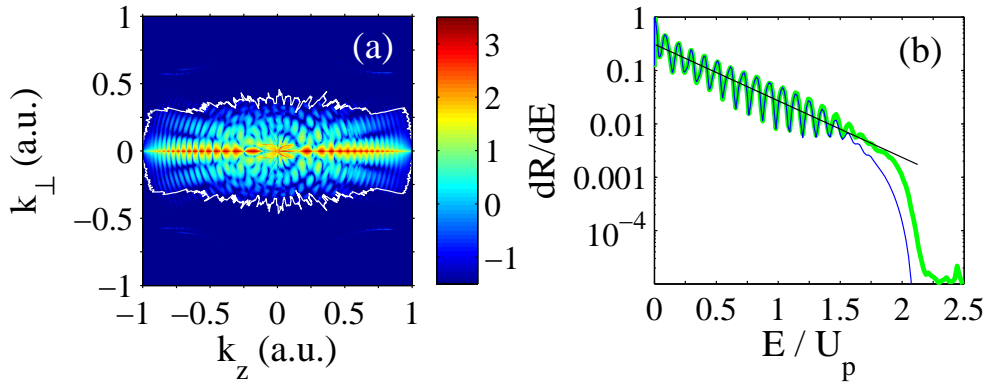} 
\end{center}
\caption{(a) The two-dimensional photoelectron momentum distribution [Eq.~(\ref{prob1})] for the Mg atom ionized by a laser pulse with an intensity of $3.0\times10^{13}$ W/cm$^2$, wavelength of 1600 nm, and duration of $n=8$ cycles. The white curve shows the boundary of the domain that can be reliably calculated using the TIPIS model. The laser field is linearly polarized along the $z$ axis. The distribution is normalized to the total ionization yield. A logarithmic color scale in arbitrary units is used. (b) Electron energy spectra calculated without any restriction on the electron trajectories [thick (green)] curve and with the exclusion of the trajectories that approach to the parent ion to the distances less than 5.0 a.u. [thin (blue) curve]. The slope of the spectra is qualitatively shown by the thin black line.} 
\label{fig1}
\end{figure} 

In Figs.~2~(a) and (b) we present the two-dimensional photoelectron momentum distributions in the $\left(k_z,k_{\perp}\right)$ plane calculated within the semiclassical model accounting for laser and Coulomb field only [panels (a) and (c)] and with account of the ME potential [panels (b) and (d)]. The first and the second row of Fig.~2, i.e., panels [(a), (b)] and [(c), (d)] show the results for Mg and Ca, respectively. The size of the bin and the number of trajectories are the same as for Fig.~1. Careful analysis of the results shown in Fig.~2 reveals that the presence of the ME term leads to two different effects: a narrowing of the longitudinal momentum distributions and a modification of the interference patterns. As we shall describe in detail later, the ME-induced dipole potential can alter the fanlike interference structures in the momentum distributions at low energy [Fig.~7~(c) and (d)].

\begin{figure}[h]
\begin{center}
\includegraphics[width=.8\textwidth]{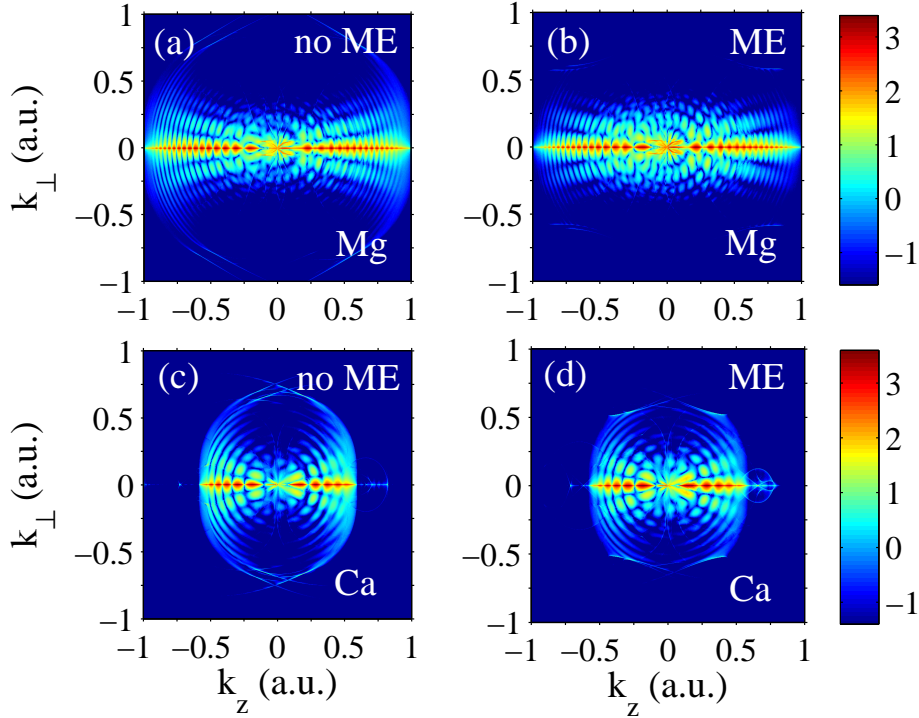} 
\end{center}
\caption{The two-dimensional photoelecton momentum distributions for Mg [(a), (b)] and Ca [(c), (d)] ionized by a laser pulse with a duration of $n=8$ cycles at a wavelength of 1600 nm. Panels (a,b) and (c,d) correspond to the intensities $3.0\times10^{13}$ W/cm$^2$ and $1.0\times10^{13}$ W/cm$^2$, respectively, implying the Keldysh parameters 0.71 and 1.13. The left column [panels (a) and (c)] show the distributions calculated ignoring the ME terms in Eqs.~(\ref{mepot}), (\ref{exiteta}), and (\ref{Phi_our}). The right column [panels (b) and (d)] displays the distributions obtained with account of the ME terms in all equations.  The distributions are normalized to the total ionization yield. A logarithmic color scale in arbitrary units is used. The laser field is linearly polarized along the z axis.}   
\label{fig2}
\end{figure} 

We first consider the narrowing effect. In order to illustrate this effect, we calculate the longitudinal momentum distributions $dR/dk_z$ with and without the ME term [see Figs.~3~(a) and (c)]. The widths of the longitudinal momentum distributions are insensitive to the interference terms, which are, therefore, not included in Figs.~3~(a) and (c). Furthermore, the narrowing of the two-dimensional distributions leads to the change of the slope of the electron energy spectra [Fig~1~(b)]. The spectra calculated with account of the ME term fall off more rapidly with electron energy than the ones calculated neglecting the polarization effects [see Figs.~3~(b) and (d)]. It is also seen from Figs.~3~(a) and (c) that the account of the ME term leads to partial filling of the dip at zero longitudinal momentum. 

\begin{figure}[h]
\begin{center}
\includegraphics[width=.8\textwidth]{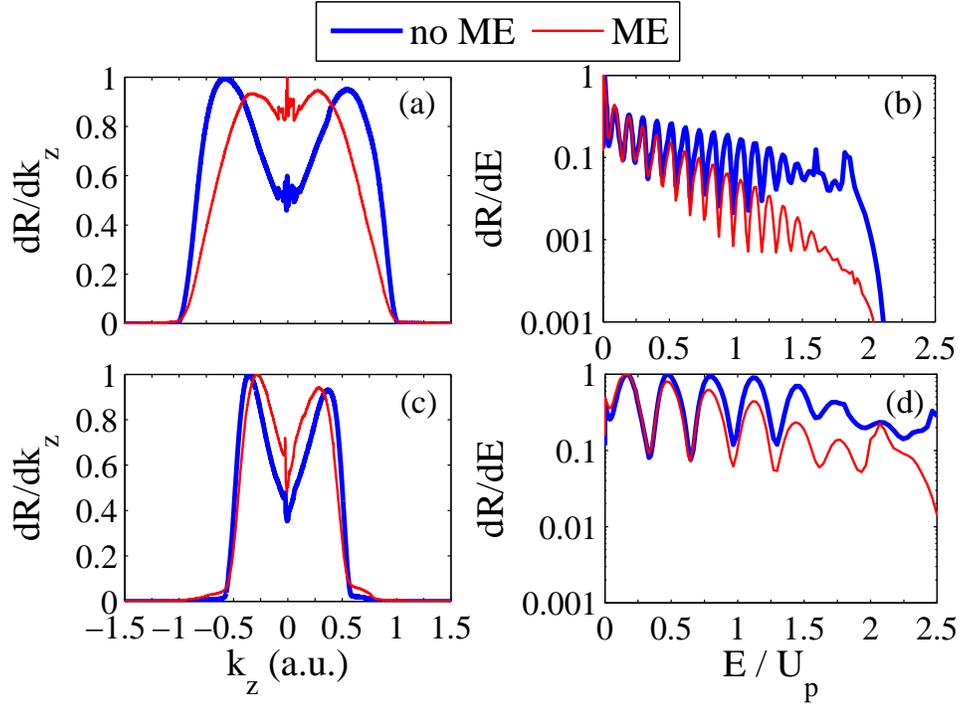} 
\end{center}
\caption{Longitudinal momentum distributions [(a),(c)] and energy spectra [(b),(d)] of the photoelectrons for ionization of Mg [panels (a) and (b)] and Ca [panels (c) and (d)]. The panels [(a),(b)] and [(c),(d)] correspond to the intensities $3.0\times10^{13}$ W/cm$^2$ and $1.0\times10^{13}$ W/cm$^2$, respectively. The wavelength and pulse durations are as in Figs.~1 and 2. The energy spectra and the longitudinal distributions are normalized to the peak value.}
\label{fig3}
\end{figure}  

In order to understand the mechanism responsible for the narrowing of the longitudinal momentum distributions, we analyze electron trajectories ending up in a bin centered at some final momentum $\vec{k}=\left(k_z,k_{\perp}\right)$. We consider ionization of Mg [see Figs~2.~(a) and (b)], and we choose $\vec{k}$ to be equal to $\vec{k}_{0}=\left(0.86, 0.31\right)$ a.u. In the importance sampling approach, where the weight of every trajectory is accounted for already at the sampling stage and the photoelectron distribution is given by Eq.~(\ref{prob1}), the presence of the ME term reduces the number of trajectories reaching this bin by a factor of 5. Therefore, the ionization probability at $\vec{k}=\vec{k_0}$ is substantially decreased due to the polarization of the residual ion. First we consider the trajectories leading to this bin neglecting the ME effect, i.e., when the electrons move in the laser and Coulomb fields. The analysis of these trajectories shows that there are three main groups of them starting from three different domains of the $\left(t_0, v_{0,\perp}\right)$ space. We refer to these trajectories as no. 1, no. 2, and no. 3, respectively. In Fig.~4 we plot one characteristic trajectory from each group when the ME term is disregarded in Eq.~(\ref{newton}) [dashed curves]. In the same plot we show the trajectories resulting when the ME term is taken into account while the initial conditions are unchanged [solid curves]. Table~1 presents detailed quantitative information about these trajectories: Their times of start $t_0^{\left(j\right)}$, initial transverse velocities $v_{0,\perp}^{\left(j\right)}$, the tunnel exit points $z_0^{\left(j\right)}$, as well as the corresponding asymptotic momenta of the electron moving in the laser field only $\vec{k}_L^{\left(j\right)}$, in both laser and Coulomb fields $\vec{k}_{CL}^{\left(j\right)}$, and, the asymptotic momentum $\vec{k}^{\left(j\right)}$ that corresponds to the case when the entire potential of Eq.~(\ref{mepot}) is included into the equations of motion (\ref{newton}) [here $j=$ 1,2 and 3]. It is seen from Table~1 and Fig.~4 that, in contrast to the trajectories no.~1 and no.~3, trajectory no.~2 is strongly affected by the ME potential. The reason is that this trajectory has the smallest exit point (see Table~1 and inset in Fig.~4). Indeed, the force acting on the electron due to the ME polarization effect decays as $1/r^2$ with increasing $r$ [see Eq.~(\ref{mepot})]. For brevity, we call this force the ME force. It is clear that the ME force can affect the electron motion only at the initial parts of the electron trajectory close to its starting point (i.e., close to the tunnel exit). The smaller the distance to the tunnel exit, the stronger the effect of the ME force on the trajectory.  

\begin{figure}[h]
\begin{center}
\includegraphics[width=.7\textwidth]{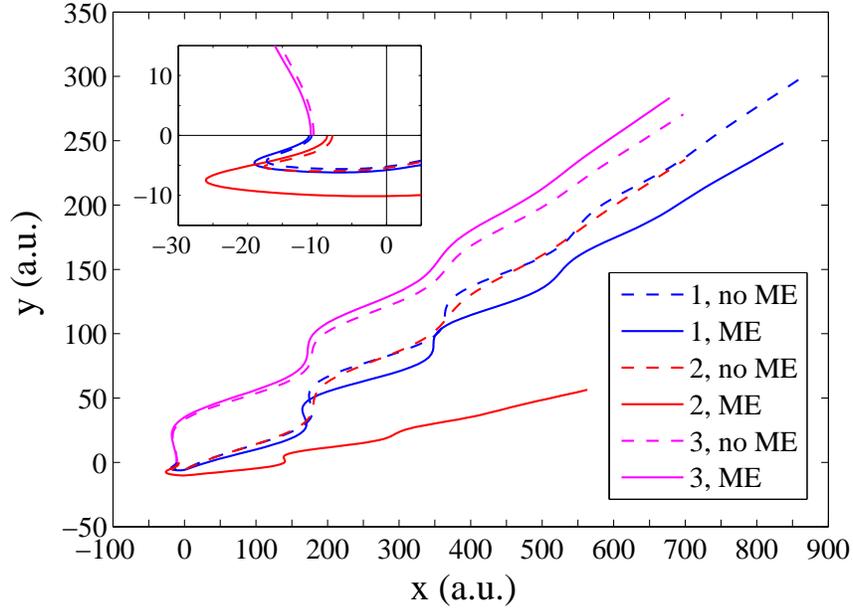} 
\end{center}
\caption{Three characteristic electron trajectories leading in the absence of the ME potential to the same final momentum $\vec{k}_{0}=\left(0.86, 0.31\right)$~a.u. The parameters correspond to the ionization of Mg by a laser pulse with a duration of $n=8$ cycles, intensity of $3.0\times10^{13}$ W/cm$^2$, and wavelength of 1600 nm. The dashed curves show the trajectories calculated ignoring the ME potential, i.e., when accounting for only the laser and Coulomb fields. The solid curves depict the trajectories moving in the laser field and the full potential of Eq.~(\ref{mepot}) including the ME term. The inset shows a zoom-in of the initial part of the electron trajectories.}   
\label{fig1}
\end{figure}

\begin{table}
\caption{\label{table1} The kinematic characteristics of the trajectories shown in Fig.~4. The table presents the times of start $\omega t_{0}^{j}$, initial transverse velocities $v_{0,\perp}^{\left(j\right)}$, starting points $z_{0}^{\left(j\right)}$, and the final asymptotic momenta $\vec{k}_{L}^{\left(j\right)}$, $\vec{k}_{CL}^{\left(j\right)}$, and $\vec{k}^{\left(j\right)}$ that correspond to the motion in the laser field only, in the laser and Coulomb fields, and in the laser field and the full potential of Eq.~(\ref{mepot}), respectively.}  
\begin{ruledtabular}
\begin{tabular}{lllllll}
$j$ &  $\omega t_{0}^{\left(j\right)}$ (rad) & $v_{0,\perp}^{\left(j\right)}$ (a.u.) & $z_{0}^{\left(j\right)}$ (a.u.) & $\vec{k}_{L}^{\left(j\right)}$ (a.u.) & $\vec{k}_{CL}^{\left(j\right)}$ (a.u.) & $\vec{k}^{\left(j\right)}$ (a.u.) \\
\hline \\
1 & 19.49 & -0.10 & -10.78 & (0.54, -0.10) & (0.84, 0.31) & (0.81, 0.26) \\
2 & 25.50 & -0.14 & -7.78 & (0.37, -0.14) & (0.86, 0.31) & (0.70, 0.09) \\
3 & 25.86 & 0.46 & -10.49 & (0.67, 0.46) & (0.86, 0.31) & (0.84, 0.32) \\
\end{tabular}
\end{ruledtabular}
\end{table}

It is seen that for trajectory no.~2 both longitudinal and transverse components of the asymptotic momentum $\vec{k}$ are reduced due to the ME force when compared to the corresponding components of the momenta $\vec{k}^{\left(1\right)}$ and $\vec{k}^{\left(3\right)}$. As the result, trajectory no.~2 will not end up in any bin of the momentum space close to $\vec{k}_{0}$. Instead, it will lead to another bin with smaller $k_z$. 
It is worth noting that for close to circularly polarized fields the ME effect manifests itself in the rotation of the two-dimensional momentum distribution towards the minor axis of the polarization ellipse \cite{Keller2012}. 

If the Coulomb and the ME forces are small compared to the laser field, these forces can be considered as small perturbations. Based on this idea analytical estimates of the effects of the Coulomb and ME forces were obtained in Ref.~\cite{Shvetsov2012} for the asymptotic electron momenta by integrating both Coulomb and ME forces along the trajectory generated by a constant field $\vec{F}\left(t_0\right)$ at the time of ionization. In linearly polarized fields these estimates may be inapplicable even for the trajectories that are not substantially affected by the ME force (e.g., trajectory no.~1). This becomes clear already from the fact that the Coulomb potential changes the sign of the transverse momentum component (cf. $\vec{k}_{C}$ and $\vec{k}$ for the trajectory no.~1). We note, however, that the sign of the ME contribution to the final electron momentum is predicted correctly by the estimates of Ref.~\cite{Shvetsov2012}.

As the narrowing of the momentum distributions due to the polarization of the ionic core is a pronounced effect, we may expect that the inclusion of the ME term will be important to explain experimental data. In Figs.~5~(a) and (b) we show the results of our semiclassical simulations for Ar ($I_{p}=0.58$~a.u., $\alpha_{I}=7.2$~a.u.) by the eight-cycle laser pulse with intensity $5.0\times10^{14}$ W/cm$^2$ and wavelength 800 nm. These parameters are close to those used in the experiment of Ref.~\cite{Rudenko2004}. Convergence was achieved at 1.6 billion trajectories and the bin size equal to $1.3\cdot10^{-3}$~a.u. Since the interference oscillations are strong and the narrowing effect is weaker for Ar than for Mg or Ca, we again neglect quantum interference when calculating the longitudinal momentum distributions [see Fig.~5~(a)]. The narrowing of the longitudinal distribution and the change of the slope of the energy spectra are clearly seen from Figs.~5~(a) and (b). The experimental photoelectron momentum distributions of Ref.~\cite{Rudenko2004} are narrower than the corresponding theoretical results based on the solution of the TDSE within the SAE (see Refs.~\cite{Morishita2007, Madsen2011}). This suggests that polarization effects may be important in resolving the remaining subtle discrepancy between the experiment \cite{Rudenko2004} and theory. 

\begin{figure}[h]
\begin{center}
\includegraphics[width=.8\textwidth]{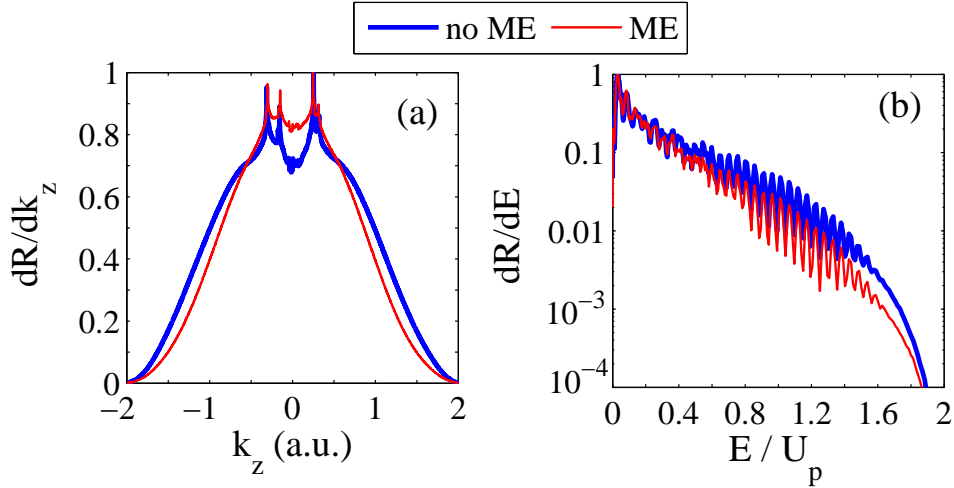} 
\end{center}
\caption{Longitudinal momentum distributions (a) and electron energy spectra (b) calculated for ionization of Ar by a Ti:sapphire laser pulse (800 nm) with a duration of 8 cycles and intensity $5.0\times10^{14}$ W/cm$^2$. Red (thin) and blue (thick) curves correspond to the semiclassical simulations with and without ME term, respectively.}    
\label{fig5}
\end{figure} 

Let us finally discuss the interference effects caused by the laser-induced polarization of the atomic residual. It is seen from Figs.~2~(a) and (b) that the changes of the interference patterns due to the ME terms in the equations of motion [Eq.~(\ref{newton})] and phase [Eq.~(\ref{Phi_our})] are not very strong. These changes are only visible in the first and partially the second ATI peaks, as well as in the vicinity of the $k_z$ axis. In order to understand the mechanism of the ME polarization-induced interference effect, we compare the photoelectron momentum distributions calculated without considering the ME effects [Fig.~6~(a)], with the account of the ME term only in the equations of motion [Fig.~6~(b)], and with the full account of the ME effects, i.e., by including the ME terms in the equations of motion and in the phase of Eq.~(\ref{Phi_our}) [Fig~6~(c)]. It is seen that the interference structures change mainly due to the presence of the ME term in the Newton's equations (\ref{newton}), i.e., due to the change of the electron trajectories caused by the polarization of the core. 

\begin{figure}[h]
\begin{center}
\includegraphics[width=.9\textwidth]{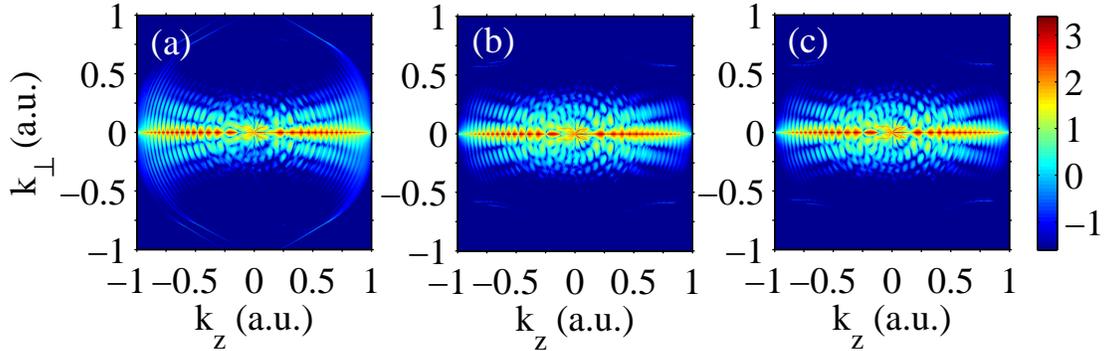} 
\end{center}
\caption{The two-dimensional photoelectron momentum distributions for ionization of Mg calculated (a) ignoring ME polarization potential, (b) accounting for the ME force in the equations of motion [Eq.~(\ref{newton})], but disregarding the ME potential in the phase [Eq.~(\ref{Phi_our})], and (c) with the full account of the ME term. The laser parameters are as in Figs.~1~(a) and (b). The distributions are normalized to the total ionization yield. A logarithmic color scale in arbitrary units is used.}   
\label{fig6}
\end{figure} 

At first glance, the facts that for Mg and Ca the polarization-induced changes of the interference patterns (i) are relatively weak, and (ii) originate due to the dynamic effect may appear counterintuitive. Indeed, due to the relatively high values of $\alpha_{I}$ for Mg and Ca, the ME term in the integrand of Eq.~(\ref{Phi_our}) seems to have a sufficiently large value to produce substantial contribution to the phases of trajectories shown in Fig.~4. Therefore, we could expect substantial modification of the interference patterns for the parameters of Figs.~2 when including the ME potential. The detailed analysis of the trajectories interfering in different bins shows, however, while the ME phases are large, they have very similar magnitudes and, hence, they do not change the interference. The reason for this is the following. In order for two trajectories to interfere maximally, they must have comparable weights. Since the tunneling probability [Eq.~(\ref{tunrate})] is a sharp function of the electric field $F\left(t_0\right)$ at the time of start, the interfering trajectories start at the time instants that correspond to similar values of the instantaneous field. Furthermore, the contribution of the ME term to the phase (\ref{Phi_our}) is mostly created on the initial part of the electron trajectory close to the tunnel exit. The latter depends only on the parameters of the atomic (molecular) species and the laser field at the time of ionization [see Eq.~(\ref{exiteta})]. As a result, the interfering trajectories have similar values of the ME contributions to the phase, $-\int_{t_0}^{\infty}\alpha_{I}\vec{F}\cdot\vec{r}/r^3 dt$, and the \textit{difference} of these contributions, which is the quantity relevant for the interference, is small. Nevertheless, for atoms and molecules with larger values of the ionic polarizability $\alpha_{I}$ this difference can reach significant values, and, therefore, produce considerable changes of the interference patterns. To illustrate this point, in Figs.~7~(a) and (b) we show the two-dimensional photoelectron momentum distributions for ionization of Ba ($\alpha_{I}=124.15$~a.u., see Ref.~\cite{Mitroy2010}) calculated without considering the ME term in the phase and with [Eq.~(\ref{Phi_our})]  account of this term, respectively. The bin size and the number of trajectories in the ensemble are the same as for Fig.~1. In order to enhance intracycle interference, we consider here a shorter pulse with a duration of $n=4$ cycles (cf. to $n=8$ in Figs.~1-6). It is seen from Figs.~7~(a) and (b) that the presence of the ME term in the phase of Eq.~(\ref{Phi_our}) leads to changes in the interference pattern. For example, the number of radial nodal lines in the fanlike interference structure for $\left|k\right|\leq 0.25$~a.u. is different in the distributions calculated without and with the ME term in the phase [cf. Figs.~7~(c) and (d)]. In Fig. 7(c), we see 6 fanlike structures for $k_\perp > 0$, while the presence of the ME contribution reduces the number of such structures to 5 in Fig.~7~(d).
\begin{figure}[h]
\begin{center}
\includegraphics[width=.75\textwidth]{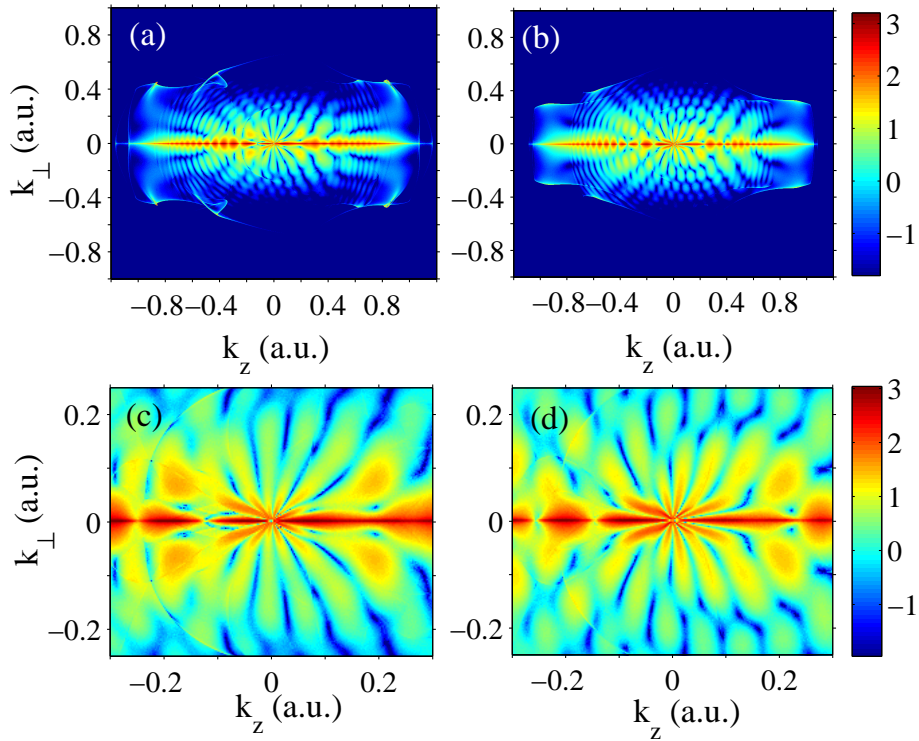} 
\end{center}
\caption{Two-dimensional electron momentum distributions for the Ba atom ionized by a laser pulse with a duration of $n=4$ cycles, wavelength of 1600 nm and an intensity of $3.0\times10^{13}$ W/cm$^2$ calculated [(a),(c)] disregarding the ME term in the phase [Eq.~(\ref{Phi_our})], and [(b),(d)] with the account of this term. Panels (c) and (d) show the magnification for $k_{z}\leq 0.3$~a.u and $k_{\perp}\leq 0.25$~a.u. of the distributions shown in (a) and (b), respectively. For both distributions the ME force is included in the equations of motion. The distributions are normalized to the total ionization yield. A logarithmic color scale in arbitrary units is used.}   
\label{fig7}
\end{figure} 
\section{Conclusions and outlook}
We have investigated ME effects as described by a laser-induced dipole polarization potential on  photoelectron momentum distributions from strong-field ionization in a linearly polarized laser field. To this end, we have applied semiclassical simulations based on the TIPIS model \cite{Keller2012}. We have analyzed the applicability of the TIPIS approach to the case of linear polarization. We have proposed a simple procedure that allows to find the domain in the photoelectron momentum distributions that can be reliably calculated by the TIPIS model. For the atomic species and laser parameters considered here this domain includes the whole direct part of the ATI spectrum. In order to study the polarization-induced interference efffects, we have combined the TIPIS approach with the SCTS model \cite{Shvetsov2016}.

We predict a pronounced narrowing of the photoelectron momentum distributions in the longitudinal direction parallel with the laser polarization. By analyzing the characteristic electron trajectories we have studied the mechanism underlying the narrowing effect. We have shown that the narrowing is caused by the polarization-induced dipole force on electrons that start relatively close to the origin.



We have also revealed the polarization-induced modification of interference effects in the photoelectron momentum distributions. This effect is found to be pronounced for atoms with relatively high static polarizabilities, and it was found to change the number of fanlike interference structures at low energy in the two dimensional electron momentum distribution. Due to the rapid progress in experimental techniques, it is now possible to study photoelectron momentum distributions with high resolution (see, e.g., Ref.~\cite{Korneev2012}), and, therefore, ME effects will have to be taken into account for accurate description of experimental data, in particular for larger molecules with large polarizabilities. 

\section{Acknowledgment} 
This work was supported by the Deutsche Forschungsgemeinschaft (Grant No.~SH~1145/1-1). Research of LBM was supported by the Villum Kann Rasmussen center of excellence, QUSCOPE - Quantum Scale Optical Processes.

\end{document}